\documentclass{aa} 
\usepackage{graphicx}
\usepackage{natbib}
\begin{document}
\title{IR photometric properties of  Red Giants in $\omega$~Cen\thanks{Based on UVES
observations collected at the European Southern Observatory,
 La Silla, Chile,
within the observing programmes 64.N-0038 and 68.D-0287.}}

\author {A. Sollima\inst{1}\fnmsep\inst{2}  
        \and
	F.R. Ferraro\inst{1}   
        \and 
         L.Origlia\inst{2}
        \and
        E. Pancino\inst{2}
        \and
        M. Bellazzini\inst{2}
        }

\offprints{A. Sollima}

\institute{Dipartimento di Astronomia, Universit\`a di Bologna, via Ranzani
           1, I-40127 Bologna, Italy\\\email{ferraro@apache.bo.astro.it;
	   antonio@omega.bo.astro.it}
           \and
            Osservatorio Astronomico di Bologna, via Ranzani 1, I-40127
           Bologna, Italy
           }

\date{Received Dec 00, 0000; accepted Dec 00, 0000}


\abstract{We present a near-infrared J and K photometric catalog
containing more than 73,000 stars in the central region of the giant
Globular Cluster $\omega$~Centauri. This is the largest IR data-set
ever published for this cluster and has been used to completely
characterize  the morphology and the properties of the Red Giant Branch
(RGB). In particular, we concentrated our attention on (i) the
anomalous RGB (RGB-a), recently discovered in this cluster and (ii) the
RGB of the dominant metal poor population (RGB-MP) in both the infrared
$(K,J-K)$ and optical-infrared $(K,V-K)$ color magnitude diagrams. The
full set of morphological parameters and photometric indices has been
measured and compared with the empirical relations by Ferraro et al.
(2000). We find that the detailed photometric properties of the RGB-a
are in full agreement with the recent spectroscopic metallicity estimates,
that place it at the metal-rich extreme of the stellar population mix
in $\omega$~Centauri.

\keywords{globular cluster: $\omega$~Cen --
                stars: Population II --
                stars: evolution --
		techniques: photometry --
		Infrared: stars  
                }                   
		}
\maketitle


\section{Introduction}

The origin and star formation history in $\omega$~Centauri, the most
luminous and massive globular cluster in our Galaxy, is one of the most
intriguing problems of modern stellar astrophysics. $\omega$~Centauri
is the only known Galactic globular cluster which shows clear
variations in the metal content of its giants. This evidence
has been firmly estabilished in the past by extensive low 
(Norris et al. 1996, Suntzeff \& Kraft 1996) and high resolution
(Norris \& Da Costa 1995, Smith et al. 2000) spectroscopic surveys.
More recently, the scenario has become more complicated due to the
discovery of an additional, metal-rich population with its own distinct
RGB (hereafter RGB-a) that contains approximately 5$\%$ of the red
giants in $\omega$~Cen (Lee et al. 1999, Pancino et al. 2000, 2002). 
In spite of the huge observational effort carried out so far, the
global picture of the cluster formation and evolution is far from being
completely understood. 

In this framework, we have started a long--term project devoted to the
detailed study of the properties of the different stellar populations
in this cluster (see the overview of the project by Ferraro, Pancino \&
Bellazzini, 2001). Within this project a number of results have been
published, in particular on the identification of the anomalous
RGB-a, and on the definition of its chemical and kinematic properties
(see Pancino et al. 2000, 2002, 2003; Bellazzini et al. 2001; Ferraro
et al. 2002; Origlia et al. 2002).
Most of the actual observational knowledge comes
from the optical (photometric or spectroscopic) study of RGB stars. 
Only a few sparse literature of near
infrared observations existed up to now. Two pioneering studies by
Glass \& Feast (1973, 1977) and Persson et al. (1980) present J, H and
K magnitudes for a few tens of bright giants. More recently, the NICMOS
camera on board of HST has been used by Pulone et al. (1998) to obtain
extremely deep photometry of a tiny area ($20"\times20"$) 7 arcminutes
away from the cluster center. The 2MASS survey has instead covered a
very wide area ($3^{\circ}\times2^{\circ}$) around $\omega$~Cen, which
however does not include the central region of the cluster. 
This paper presents a large J and K photometric
catalog of more than 73,000 stars in an area covering $\sim
13'\times13'$ around the center of $\omega$~Cen. By combining the
IR-data set with wide field optical photometry (Pancino et al. 2000,
2003), we measured the complete set of morphological parameters defined
by Ferraro et al. (2000, hereafter F00), which fully characterize the
photometric properties of the RGB of different populations in
$\omega$~Centauri. 


\section{Observational Material}

   \begin{figure}
   \centering
    
   \includegraphics[width=8.7cm]{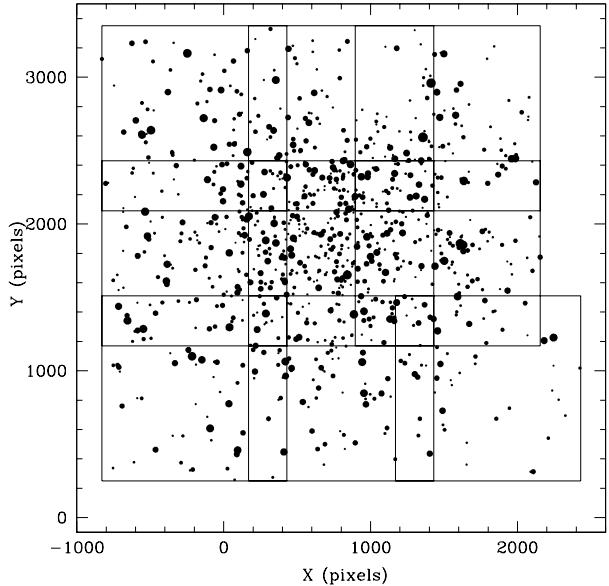}
   \caption{ Map of the region sampled by the IR observations. North is
   up, east on the left. Only stars with $K<13$  have been
   plotted.}
   \label{mosaic}%
   \end{figure}

A large set of J and K images were obtained at ESO, La silla (Chile),
during two observing runs (Run $\#1$ and $\#2$, see
Table~\ref{obs_log}) at the New Technology Telescope (NTT) using the
near-IR imager/spectrometer SOFI. The observations were secured  as
back-up programmes. SOFI is an
imager/spectrometer equipped with a $1024\times1024$ Rockwell IR-array
detector. All the observations presented here were performed with a
scale of $0.292''/pixel$, fixing a global FoV of $\sim 5'\times 5'$
each field.

During $Run\#1$, we covered the central region of the cluster with a
mosaic of nine partially overlapping fields (hereafter {\em
Science-fields}), sampling a total area of $\sim 13'\times 13'$
around the cluster center (see Figure~\ref{mosaic}). A set of high S/N
flatfields in each filter has been obtained with an halogen lamp,
alternatively switched on and off. For each field observed in the
cluster, a corresponding sky-field has been also observed ($\sim 10$
arcmin away from the cluster center) using the same instrument
configuration. Each sky-image was obtained as the median of at least 5
frames, shifted by some hundreds of pixels with respect to each other.
The final set consists of 18 sky--subtracted and flatfield corrected
images, 9 for each filter.

During $Run\#2$ a set of 4 additional fields (hereafter {\em
Calibration-fields}), entirely within the area sampled in $Run\#1$,
have been obtained for photometric calibration purposes. The same
pre-reduction procedure used for the {\em Science-fields} has been also
applied to the {\em Calibration-fields}.

\begin{table}
\caption[]{Observing logs. }
\label{obs_log}
$$ 
\begin{array}{cccc}
\hline
\noalign{\smallskip}
Run & Date & Filter & ~~~NDIT \times DIT \\
\noalign{\smallskip}
\hline
\noalign{\smallskip}
 1  & \rm{13-14~Jan~2000}   & J       & 45 \times 1.2  \\
 1  & \rm{13-14~Jan~2000}   & K_s   & 150 \times 1.2  \\
 1  &  \rm{13-14~Jan~2000}   & J       & 15 \times 1.2  \\
 1  &  \rm{13-14~Jan~2000}   & K_s   & 60 \times 1.2  \\
 2  & \rm{30-31~Dec~2001}   & J   & 60 \times 1.2  \\
2  &   \rm{30-31~Dec~2001}  & K_s   & 120 \times 1.2  \\
\noalign{\smallskip}
\hline
\end{array}
$$  
\end{table}

\begin{table}
\caption[]{A sample of the online catalog. Only a few entries are
displayed to illustrate the catalog format and contents. See text for a
detailed description on the measurement procedures.}
\label{online}
$$ 
\begin{array}{ccccccc}
\hline
\noalign{\smallskip}
Id & J & K & \sigma_J & \sigma_K & RA & Dec \\
 & (mag) & (mag) & (mag) & (mag) & (deg) & (deg) \\
\noalign{\smallskip}
\hline
\noalign{\smallskip}
1 & 8.728 & 7.731 & 0.043 & 0.047 & 201.7852918 & -47.5302560 \\
2 & 9.178 & 8.291 & 0.100 & 0.038 & 201.7787851 & -47.5318524 \\
3 & 9.134 & 8.103 & 0.063 & 0.135 & 201.7231633 & -47.5327578 \\
4 & 9.274 & 8.305 & 0.077 & 0.183 & 201.7602655 & -47.4837799 \\
5 & 9.305 & 8.401 & 0.070 & 0.059 & 201.7608180 & -47.5516698 \\
\noalign{\smallskip}
\hline
\end{array}
$$ 
\end{table}

All photometric reductions were carried out using ROMAFOT (Buonanno et
al. 1983). The details of the reduction procedure have already been
reported in previous papers (Ferraro et al. 1994, Ferraro et al. 1995).
We just mention here that for the IR--frames we used the PSF--fitting
routine specifically modified to deal with under-sampled stellar images
(Buonanno \& Iannicola, 1988). The source detection was performed
independently on each field; more than 80,000 separate stars have
been detected in total in the nine {\rm science fields} observed during
$Run \#1$. 

\subsection{Photometric Calibration}
\label{phot}

   \begin{figure}
   \centering
    
   \includegraphics[width=8.7cm]{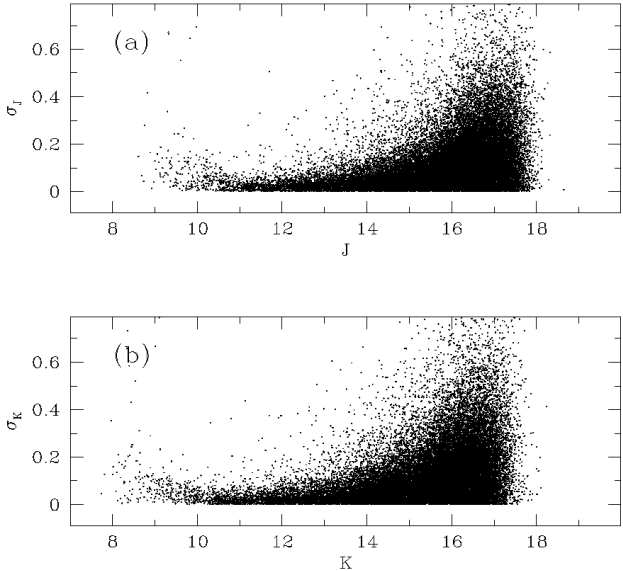}
   \caption{J ({\it panel} a) and K ({\it panel} b) photometric errors as a
   function of magnitude, represented by the standard deviation from the
   mean of repeated measurements.}
   \label{err}%
   \end{figure}

During $Run \#2$, a set of nine standard stars from the  Persson et al.
(1998) list has been observed together with the 4 {\em
Calibration-fields} in the cluster. Five measurements for each standard
have been secured and averaged, for each filter. The calibrating
equations linking the aperture photometry to the standard system are:
$$ K = k + 22.30 \pm 0.02$$ $$ J = j + 23.03 \pm 0.02$$ where $j$ and
$k$ are the instrumental magnitudes, $J$ and $K$ the calibrated ones.
The existence of a slight residual colour equation (with slope~$<0.01$) in
both filters cannot be totally excluded.

The two above equations have been used to calibrate the 4 {\em
Calibration-fields} in $\omega$ Cen: as usual the most isolated and
bright stars in the field have been employed to link the aperture
magnitudes to the fitting instrumental ones, after normalizing for exposure time
and correcting for airmass. A catalog with more than 10,000 calibrated
 stars in the cluster has been produced that has
been used to calibrate the global IR catalog, including
more than 73,000 stars. 

Our calibration has been compared with the previous photometric catalogues.
Unfortunately, only 8 stars were found in common between the present
catalog and the pioneering work by Persson et al. (1980), while no star
are common with the very deep study by Pulone et al.
(1998). The average magnitude differences found are $\Delta J = 0.034
\pm 0.063$ and $\Delta K = 0.093 \pm 0.079$ ,fully consistent 
with no systematic offset between our catalog and the one by 
Persson et al. (1980) within the errors. A much more
significant comparison can be achieved with the 2MASS catalog \footnote{See 
Cutri et al. (2003), Explanatory Supplement to the 2MASS All Sky Data Relase,
http://www.ipac.caltech.edu/2mass/ } (see
also Section 5), showing that our calibrated magnitudes are in
very good agreement with the 2MASS photometric system ($\Delta
J=0.00\pm0.10$ and $\Delta K=-0.04\pm0.10$). 

\subsection{Astrometry and Resulting Catalog}

   \begin{figure}
   \centering
    
   \includegraphics[width=8.7cm]{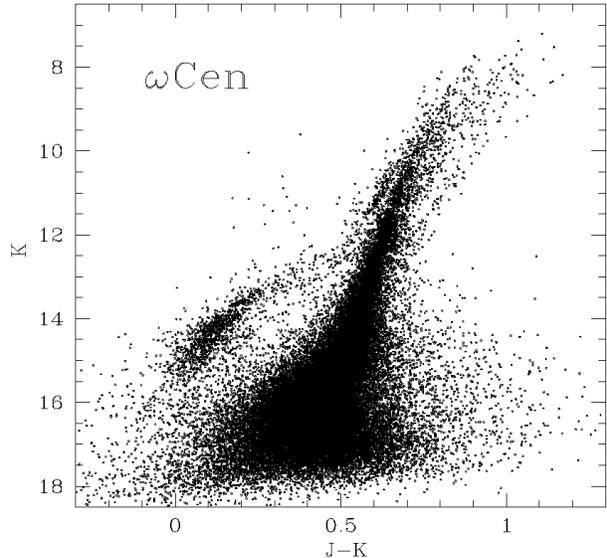}
   \caption{ $(K,J-K)$ color magnitude diagram for $\sim$ 73,000 stars
   measured in $\omega$ Cen.}
   \label{cmdjk}%
   \end{figure}

   \begin{figure}
   \centering
    
   \includegraphics[width=8.7cm]{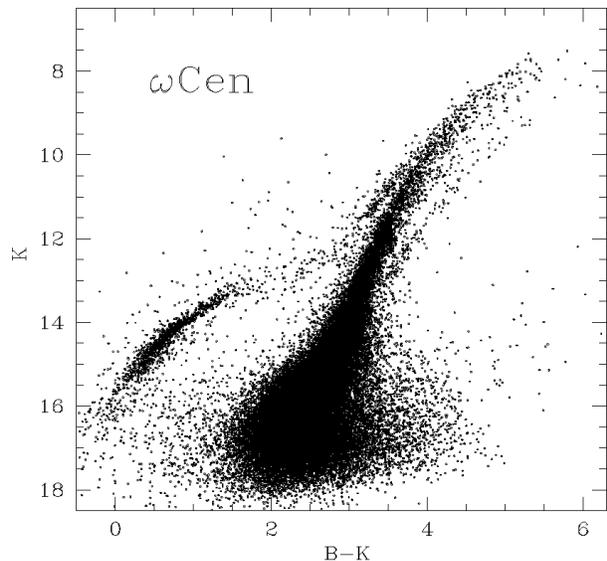}
   \caption{The combined (K,B-K) CMD for stars detected in $\omega$
   Cen. The B magnitudes are from Pancino et al. (2000).}
   \label{cmdbk}%
   \end{figure}

The photometrically calibrated catalog has been compared with the large
proper motion one by van Leeuwen et al. (2000), that lists accurate
positions for more than 10,000 red giants in $\omega$~Cen. We used a
software package under development at the Bologna Observatory
(Montegriffo et al. 2003, in preparation), specifically designed to
correlate populous and crowded stellar catalogs, and to obtain precise
astrometry of large fields. We thus were able to produce coordinates in
the J2000 reference frame for stars in the whole catalog, with a
typical r.m.s of 200 mas. 

The final on-line catalog format and content is illustrated in
Table 2. For each star we present, along with the star
number and coordinates, the J and K magnitudes and their errors,
obtained as the standard deviation from the mean of repeated
measurements: when a star was measured only once, we assigned a null
error value to its magnitude.  Figure~\ref{err} shows the typical
photometric errors for the J and K bands, as a function of the magnitude.
The mean error at the Horizontal Branch (HB) magnitude level ($ J
\simeq 14.2, K \simeq 14$) is about 0.02 mag in both bands.
Figure~\ref{cmdjk} presents the $(K,J-K)$ color Magnitude Diagram (CMD)
obtained from the final IR
catalog.                                            

The IR catalog has also been correlated with the partially published
wide field optical catalog by Pancino et al. (2000, 2003). A final
multi-wavelength catalog (B, V, I, J and K) has been obtained and will
be used throughout the paper. As an example of combined
optical-infrared CMD, Figure~\ref{cmdbk} shows the $(K,B-K)$ CMD. 

   \begin{figure}
   \centering
    
   \includegraphics[width=8.7cm]{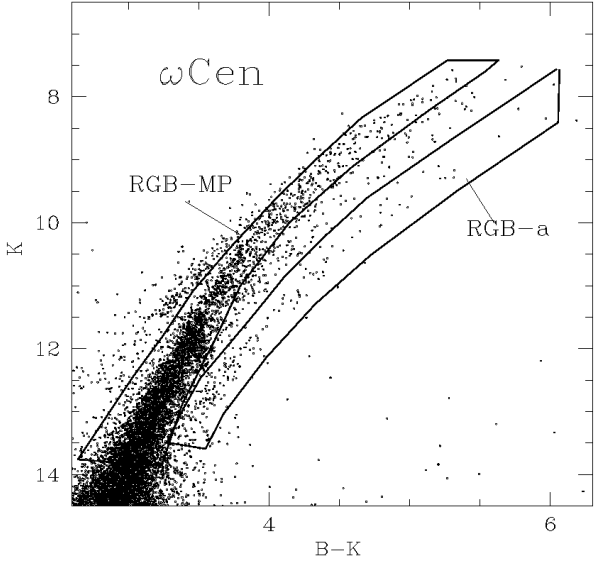}
   \caption{The zoomed CMD of the RGB region. Selection boxes for the
   metal poor (RGB-MP) and anomalous (RGB-a) population are shown.}
   \label{rgb}%
   \end{figure}

\subsection{Colour Magnitude Diagrams}

The main features of the CMDs presented in Figures~\ref{cmdjk} and
\ref{cmdbk} are schematically listed below:

\begin{enumerate}

\item{The CMDs sample the entire evolved populations in the cluster,
reaching the TO region ($K\sim 17$).} 

\item{The complex structure of the RGB in $\omega$ Cen is clearly
visible, starting from the blue side, where the
dominant metal-poor population defines the main RGB
(RGB-MP).} 

\item{The anomalous RGB (RGB-a, as defined by Pancino et al. 2000), is
clearly visible as a well separated population at the extreme red side
of the main RGB structure, having $K<13$ and $(J-K)>0.6$
(see Figure~\ref{cmdjk}) and $(B-K)>3.4$ (see Figure~\ref{cmdbk}).}

\item{An intermediate RGB component (RGB-MInt), which uniformly
populates the CMD region between the RGB-MP and the RGB-a, can also
be easily noticed.}

\item{The AGB stars appear well separated from the RGB in both CMDs.}

\item{The HB is well defined ($(B-K)<2$); in
particular, a significant population of under-luminous HB stars (D'Cruz
1999) is visible at blue colours ($(B-K)<1$), lying below the main
HB locus.}

\end{enumerate}

The dominant RGB-MP component is extremely well populated, allowing a
robust measurement of the RGB-tip (Bellazzini et al. 2003, in preparation) 
and the RGB bump position. In the following, we will
discuss the photometric properties and the evolutionary features of the
two best defined RGB populations: the RGB-MP and the RGB-a, since the
intermediate population is spreaded within these two extremes. 

 
\section{RGB fiducial lines}
\label{samples}

\begin{table}
\caption[]{Mean ridge line of the RGB-MP in $\omega$ Cen. }
\label{mrl_mp}
$$
\begin{array}{ccccc}
\hline
\noalign{\smallskip}
K & J-K & V-K & B-K & B-J \\
\noalign{\smallskip}
\hline
\noalign{\smallskip}
 7.8   &  0.90  &  3.74  &  5.24  &  4.26   \\
 8.0   &  0.88  &  3.63  &  5.09  &  4.14   \\
 8.2   &  0.86  &  3.54  &  4.96  &  4.03   \\ 
 8.4   &  0.85  &  3.45  &  4.82  &  3.93   \\ 
 8.6   &  0.83  &  3.37  &  4.70  &  3.83   \\ 
 8.8   &  0.81  &  3.29  &  4.58  &  3.73   \\ 
 9.0   &  0.80  &  3.21  &  4.47  &  3.64   \\ 
 9.2   &  0.78  &  3.14  &  4.36  &  3.56   \\ 
 9.4   &  0.77  &  3.08  &  4.27  &  3.48   \\ 
 9.6   &  0.76  &  3.02  &  4.17  &  3.40   \\ 
 9.8   &  0.74  &  2.96  &  4.08  &  3.34   \\ 
 10.0  &  0.72  &  2.90  &  4.00  &  3.26   \\ 
 10.2  &  0.72  &  2.85  &  3.92  &  3.20   \\ 
 10.4  &  0.71  &  2.82  &  3.85  &  3.14   \\ 
 10.6  &  0.70  &  2.76  &  3.78  &  3.08   \\ 
 10.8  &  0.69  &  2.72  &  3.71  &  3.02   \\ 
 11.0  &  0.68  &  2.68  &  3.65  &  2.97   \\ 
 11.2  &  0.67  &  2.64  &  3.60  &  2.92   \\ 
 11.4  &  0.66  &  2.60  &  3.54  &  2.88   \\ 
 11.6  &  0.65  &  2.57  &  3.49  &  2.83   \\ 
 11.8  &  0.64  &  2.54  &  3.44  &  2.79   \\ 
 12.0  &  0.63  &  2.51  &  3.39  &  2.75   \\ 
 12.2  &  0.62  &  2.48  &  3.35  &  2.71   \\ 
 12.4  &  0.61  &  2.45  &  3.31  &  2.68   \\ 
 12.6  &  0.61  &  2.42  &  3.27  &  2.64   \\ 
 12.8  &  0.60  &  2.39  &  3.23  &  2.61   \\ 
 13.0  &  0.59  &  2.36  &  3.19  &  2.58   \\ 
 13.2  &  0.58  &  2.33  &  3.15  &  2.54   \\ 
 13.4  &  0.58  &  2.30  &  3.11  &  2.51   \\ 
 13.6  &  0.57  &  2.27  &  3.08  &  2.48   \\ 
 13.8  &  0.56  &  2.24  &  3.04  &  2.45   \\ 
 14.0  &  0.56  &  2.20  &  3.00  &  2.42   \\ 
 14.2  &  0.55  &  2.17  &  2.96  &  2.39   \\ 
 14.4  &  0.54  &  2.13  &  2.93  &  2.36   \\ 
 14.6  &  0.54  &  2.09  &  2.89  &  2.33   \\ 
 
\noalign{\smallskip}
\hline
\end{array}
$$ 
\end{table}

\begin{table}
\caption[]{Mean ridge line of the RGB-a in $\omega$ Cen. }
\label{mrl_a}
$$ 
\begin{array}{ccccc}
\hline
\noalign{\smallskip}
K & J-K & V-K & B-K & B-J \\
\noalign{\smallskip}
\hline
\noalign{\smallskip}
    8.2  &     1.15 & 4.49 & 6.00 & 5.10 \\
    8.4  &     1.12 & 4.34 & 5.84 & 4.92 \\
    8.6  &     1.09 & 4.20 & 5.53 & 4.59 \\
    8.8  &     1.06 & 4.06 & 5.53 & 4.59 \\
    9.0  &     1.03 & 3.94 & 5.39 & 4.42 \\
    9.2  &     1.01 & 3.82 & 5.25 & 4.30 \\
    9.4  &     0.98 & 3.72 & 5.12 & 4.18 \\
    9.6  &     0.96 & 3.61 & 5.00 & 4.06 \\
    9.8  &     0.93 & 3.52 & 4.87 & 3.94 \\
   10.0  &     0.91 & 3.43 & 4.76 & 3.84 \\
   10.2  &     0.89 & 3.35 & 4.65 & 3.74 \\
   10.4  &     0.87 & 3.27 & 4.54 & 3.65 \\
   10.6  &     0.85 & 3.20 & 4.44 & 3.56 \\
   10.8  &     0.83 & 3.13 & 4.34 & 3.48 \\
   11.0  &     0.81 & 3.07 & 4.24 & 3.41 \\
   11.2  &     0.79 & 3.01 & 4.15 & 3.33 \\
   11.4  &     0.78 & 2.95 & 4.06 & 3.27 \\
   11.6  &     0.76 & 2.90 & 3.98 & 3.20 \\
   11.8  &     0.74 & 2.84 & 3.90 & 3.14 \\
   12.0  &     0.73 & 2.79 & 3.82 & 3.08 \\
   12.2  &     0.71 & 2.74 & 3.74 & 3.02 \\
   12.4  &     0.70 & 2.69 & 3.67 & 2.96 \\
   12.6  &     0.68 & 2.64 & 3.60 & 2.90 \\
   12.8  &     0.67 & 2.59 & 3.53 & 2.84 \\
   13.0  &     0.65 & 2.54 & 3.46 & 2.78 \\
   13.2  &     0.64 & 2.49 & 3.40 & 2.72 \\
   13.4  &     0.63 & 2.43 & 3.34 & 2.65 \\
   13.6  &     0.61 & 2.37 & 3.27 & 2.59 \\
   
\noalign{\smallskip}
\hline
\end{array}
$$ 
\end{table}

The first step in order to study the photometric properties of the two
populations, is the determination of the mean ridge
lines. To do this, we applied the standard procedure, already described
in Ferraro at al. (1999, 2000). 

As usual, a first selection was performed by eye. In particular, in the
case of RGB-MP, we took special care in excluding HB and AGB stars,
which are easily identified from the high quality CMDs shown in
Figure~\ref{cmdjk} and \ref{cmdbk}.  
The selection of the RGB samples could be in principle disturbed by the 
presence of the RGB-MInt stars.
However, RGB-a stars are well separated from the remaining populations
when selected from the optical-infrared CMD of Figure~\ref{cmdbk}.
The contamination of the RGB-MP population from the RGB-MInt stars can be 
quantified in $<10\%$ , with negligible impact on the ridge line determination of the former.
Figure~\ref{rgb} shows the adopted
selection boxes for the RGBs of the different populations.
We used a low-order polynomial to fit the selected stars and an
iterative procedure to automatically reject stars lying more then $2
\sigma$ away from the best-fit line. The iteration was carried on until
convergence to a stable fit was obtained. The
resulting mean ridge lines in various planes are reported in
Table 3 (for the RGB-MP) and Table 4 (for the
RGB-a).

In order to convert the derived RGB mean ridge lines to the absolute
$M_K,(J-K)_0$ and $M_K,(V-K)_0$ planes and to measure the photometric indices and
morphology parameters defined by F00 (see Section 4), we need to
make some basic assumptions for the metallicity, reddening and distance
for the sub-populations of $\omega$~Cen. 

\subsection{Metallicity}

For the dominant metal-poor (RGB-MP) population we adopted the
metallicity obtained by the most extensive spectroscopic survey
performed in $\omega$~Cen by Norris et al. 1996, who provided a peak value of 
$[Ca/H]=-1.4\pm0.1$. By using the stars in our sample which have both [Ca/H] from Norris et al. 1996
and [Fe/H] aboundances from Suntzeff \& Kraft 1996, we translated the [Ca/H] peak assumed above  
into [Fe/H] $\sim -1.6 \pm 0.1$ .
Both high-resolution optical spectra (Pancino et
al., 2002) and medium-resolution IR spectra (Origlia et al., 2003) suggest
a significantly higher metallicity for the anomalous (RGB-a) population,
[Fe/H]$=-0.6\pm0.15$.

Straniero \& Chieffi (1991) and Salaris, Chieffi \& Straniero (1993)
showed that, when computing the isochrones of Population II stars, the
contribution of the $\alpha$-element enhancement can be taken into
account by simply rescaling standard models to the global metallicity
[M/H], according to the following relation $$ \mathrm{[M/H]} =
\mathrm{[Fe/H]}+\log_{10}(0.638f_{\alpha}+0.362) $$ where $f_{\alpha}$
is the enhancement factor for the $\alpha$-elements. We assumed
[$\alpha$/Fe]$\simeq+0.3$ for the RGB-MP (Norris \& Da Costa 1995,
Smith et al. 2000) and a lower enhancement [$\alpha$/Fe]$\simeq+0.1$
for the RGB-a, according to Pancino et al. (2002) and Origlia et al.
(2003). Therefore, we obtained a global metallicity of
[M/H]$=-1.39\pm0.10$ and [M/H]$=-0.53\pm0.21$ for the RGB-MP and the
RGB-a, respectively.

   \begin{figure}
   \centering
    
   \includegraphics[width=8.7cm]{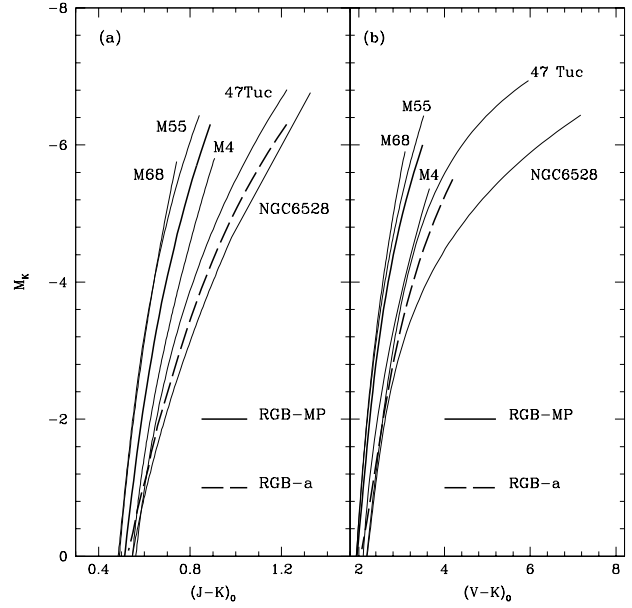}
   \caption{Fiducial ridge lines in the $M_K$,$(J-K)_0$ plane ({\it panel
   a}) and in the $M_K$,$(V-K)_0$ plane ({\it panel b}) for a few GGCs of 
   F00 and for the two populations of $\omega$~Cen.}
   \label{lines}%
   \end{figure}

\subsection{ZAHB level}

To convert the fiducial ridge lines into the absolute plane, we need to
assume a distance modulus and a reddening correction. In the following,
we adopt the distance scale presented by Ferraro et al. (1999), which
is based on the comparison between the actual level of the Zero Age 
Horizontal Branch (ZAHB) and the theoretical models computed by 
Straniero, Chieffi \& Limongi (1997). By using the optical photometric
 catalog partially published by Pancino et al. (2000) and following the 
 prescriptions of Ferraro et al. (1999), we obtain for $\omega$~Cen
$V_{ZAHB}=14.55\pm0.09$. Assuming a metallicity of $[Fe/H]=-1.6$ for
the main population (see above) and following eq. [4] by Ferraro et al.
(1999), we obtain $M_{V}^{ZAHB} = 0.56\pm0.09$. Adopting $E(B-V)=0.11
\pm 0.01 $ (Lub, 2002), the distance modulus finally turns out to be
$(m-M)_0 =13.65 \pm 0.13$, in nice agreement with the most recent
determinations by Thompson et al. (2001) and Caputo, Degl'Innocenti \&
Marconi (2002). 

\begin{table}
\caption[]{Inferred $(J-K)_0$ RGB colors at fixed magnitudes. }
\label{measjk}
$$ 
\begin{array}{ccccc}
\hline
\noalign{\smallskip}
 & (J-K)_{0}^{-5.5} & (J-K)_{0}^{-5} & (J-K)_{0}^{-4} & (J-K)_{0}^{-3} \\
\noalign{\smallskip}
\hline
\noalign{\smallskip}
RGB-MP & 0.81\pm0.02 & 0.77\pm 0.02 & 0.69\pm 0.02 & 0.64\pm 0.02 \\
RGB-a  & 1.10\pm0.03 & 1.02\pm 0.03 & 0.89\pm 0.04 & 0.79\pm 0.03 \\ 
\noalign{\smallskip}
\hline
\end{array}
$$ 
\caption[]{Inferred $(V-K)_0$ RGB colors at fixed magnitudes. }
\label{measvk}
$$ 
\begin{array}{ccccc}
\hline
\noalign{\smallskip}
 & (V-K)_{0}^{-5.5} & (V-K)_{0}^{-5} & (V-K)_{0}^{-4} & (V-K)_{0}^{-3} \\
\noalign{\smallskip}
\hline
\noalign{\smallskip}
RGB-MP & 3.24\pm0.03 & 3.03\pm 0.03 & 2.69\pm 0.03 & 2.44\pm 0.03 \\
RGB-a  & 4.19\pm0.21 & 3.83\pm 0.20 & 3.27\pm 0.09 & 2.87\pm 0.08 \\ 
\noalign{\smallskip}
\hline
\end{array}
$$ 
\end{table}
\begin{table}
\caption[]{RGB magnitudes and slopes for the RGB-MP and RGB-a of
$\omega$~Cen.}
\label{slope_tab}
$$ 
\begin{array}{cccc}
\hline
\noalign{\smallskip}
 & M_{K}^{(J-K)_0=0.7} & M_{K}^{(V-K)_0=3} & Slope_{RGB} \\
\noalign{\smallskip}
\hline
\noalign{\smallskip}
RGB-MP & -4.00\pm0.14 & -4.93\pm 0.14 & -0.050\pm 0.003 \\
RGB-a  & -2.07\pm0.16 & -3.34\pm 0.24 & -0.093\pm 0.006 \\ 
\noalign{\smallskip}
\hline
\end{array}
$$ 
\end{table}

\subsection{Comparison with the F00 sample}

The ridge lines of Tables 3 and 4 have been
transformed to the absolute planes $M_K$,$(J-K)_0$ and $M_K$,$(V-K)_0$,
using the distance modulus and the reddening discussed above, and assuming
$A_J=0.87$ $E(B-V)$ and $A_K=0.38$ $E(B-V)$ extinction coefficients  
in the J and K bands, respectively (Savage \& Mathis,1979).
 Figure~\ref{lines} compares the ridge lines of $\omega$ Cen with the lines
of a few globular clusters belonging to the sample presented by F00. 

In both planes the RGB-MP ridge line is located between the metal-poor
clusters M~68 ($[Fe/H] \sim -1.99$) and M~4 ($[Fe/H] \sim -1.19$), close to M55 
($[Fe/H] \sim -1.61$) as expected, given its metallicity.
The RGB-a is located between the metal-rich clusters 47 Tuc ($[Fe/H] \sim -0.7$) 
and NGC~6528 ($[Fe/H] \sim -0.38$), in good agreement with the spectroscopic 
constrains.


\section{RGB colors and morphology}
\label{param}

   \begin{figure}
   \centering
    
   \includegraphics[width=8.7cm]{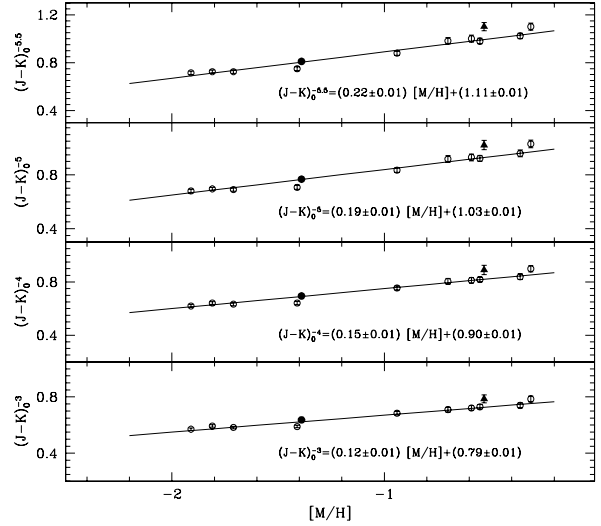}
   \caption{ RGB mean $(J-K)_0$ colors at different magnitudes ($M_K
   =-3,-4,-5,-5.5$) as a function of global metallicity for the ten
   GGCs of F00 (open circles), the RGB-MP (filled circle) and the RGB-a
   (filled triangle).}
   \label{coljkmh}%
   \end{figure}

   \begin{figure}
   \centering
    
   \includegraphics[width=8.7cm]{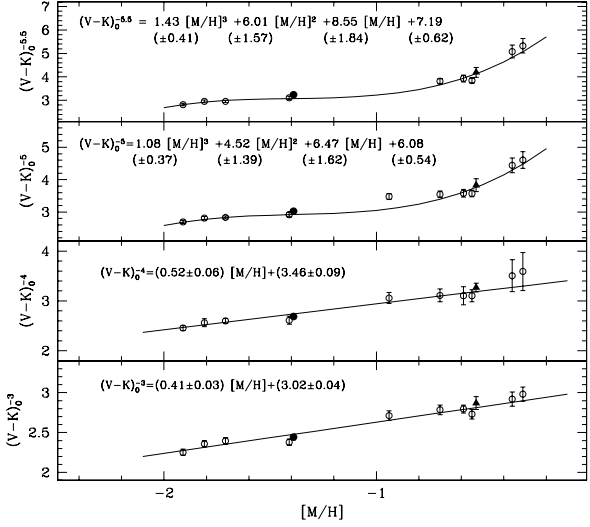}
   \caption{ RGB mean $(V-K)_0$ colors at different magnitudes ($M_K
   =-3,-4,-5,-5.5$) as a function of global metallicity for the ten
   GGCs of F00 (open circles), the RGB-MP (filled circle) and the RGB-a
   (filled triangle).}
   \label{colvkmh}%
   \end{figure}

In order to quantitatively describe the morphology and properties of
the RGBs of the different populations in $\omega$ Cen, we measured the
full set of photometric parameters defined by F00. 

The main RGB observables measured in the IR CMDs are the following:

\begin{itemize}

\item{The RGB $(J-K)_0$ and $(V-K)_0$ colors at different magnitude
levels;} 

\item{The RGB $M_K$ magnitude at fixed colors;} 
    
\item{The RGB slope.}
  
\end{itemize}
 
In Tables~\ref{measjk} and \ref{measvk}, the inferred $(J-K)_0$ and
$(V-K)_0$ colors at fixed $M_K$ magnitudes are listed. Our results
 are in good agreement with the empirical relations proposed by F00 in both
colors and magnitudes, the $(J-K)_0$ colors of
the RGB-a ridge line being only slightly redder than predicted, but still within
the overall uncertainties. 

   \begin{figure}
   \centering
    
   \includegraphics[width=8.7cm]{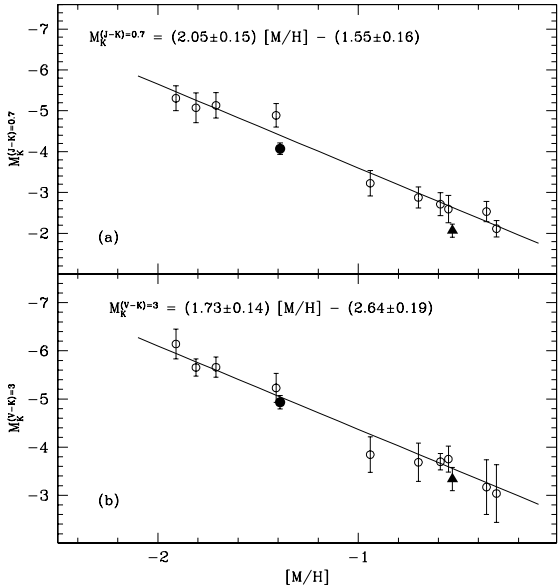}
   \caption{ $M_K$ at constant $(J-K)_0=0.7$ ({\it panel} a) and $(V-K)_0=3$
   ({\it panel} b) as a function of  global metallicity for the ten GGCs of
   F00 (open circles), the RGB-MP (filled circle) and the RGB-a (filled
   triangle).}
   \label{magk}%
   \end{figure}

   \begin{figure}
   \centering
    
   \includegraphics[width=8.7cm]{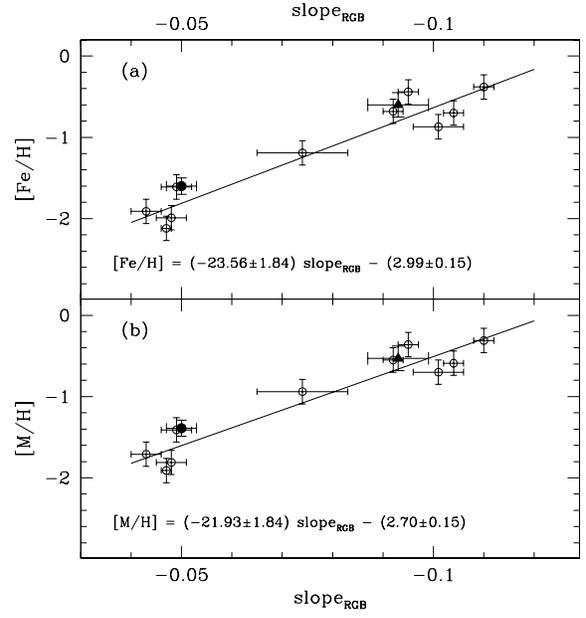}
   \caption{Metallicity scales: [Fe/H] ({\it panel} a) and [M/H] 
   ({\it panel} b) as a function of the derived RGB slope for the ten 
   GGCs of F00 (open circles), the RGB-MP (filled circle) and the RGB-a 
   (filled triangle).}
   \label{slope}%
   \end{figure}
 
Following F00, we also measured the absolute $M_K$ magnitude at
$(V-K)_0=3$ and $(J-K)_0=0.7$. In figure~\ref{magk} the dependence of
these two parameters on global metallicity is shown; the measurements
taken on the RGB-MP and RGB-a mean lines of $\omega$ Cen are
overplotted as solid symbols. At $(V-K)_0=3$, we observe a nice
agreement between the magnitudes measured for both the RGB-MP and RGB-a
populations and the relations proposed by F00. At $(J-K)_0=0.7$,
the absolute $M_K$ magnitudes observed for the two RGB
populations of $\omega$ Cen are instead clearly underestimated. This
discrepancy can be due to uncertaintes in the reddening assumptions: in
fact errors of a few hundredths of magnitude produce uncertainties of
about 0.2-0.3 in K, depending on the RGB region intercepted. 

To further describe the properties of the RGBs of $\omega$ Cen, we have
measured the RGB slope, adopting the technique described by Kuchinski
et al. (1995) and using the RGB samples described in
Section 3 and shown in Figure~\ref{rgb}. Following the
prescriptions by Kuchinski et al. (1995), only the brightest portion of
the RGB ($\sim$1 mag above the HB level) has been considered in the
fit. As already emphasized by F00, although linear fits are not the
best representation of the RGB shape, the measurement of the RGB slope
is still important since it represents a distance and reddening
independent parameter to describe the RGB morphology. The inferred
slopes for the RGB-MP and the RGB-a are consistent, within the
uncertainties, with those of clusters of similar metal content
(see Figure~\ref{slope}). The adopted absolute $M_K$ magnitudes at
$(V-K)_0=3$ and $(J-K)_0=0.7$ and the slopes of the RGB for the two
main populations of $\omega$ Cen are listed in Tables 7.

   \begin{figure}
   \centering
    
   \includegraphics[width=8.7cm]{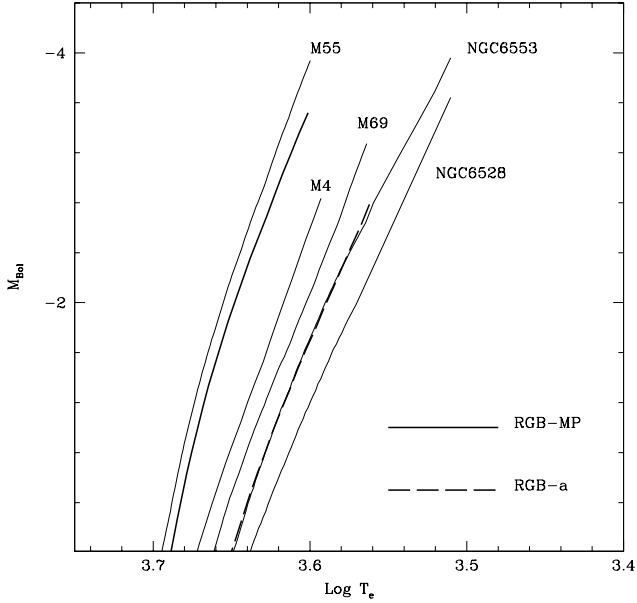}
   \caption{RGB fiducial ridge lines in the $M_{Bol},\log T_e$
   theoretical plane for five GGCs of F00 and for the two populations
   of $\omega$ Cen. }
   \label{teo}%
   \end{figure}

   \begin{figure}
   \centering
    
   \includegraphics[width=8.7cm]{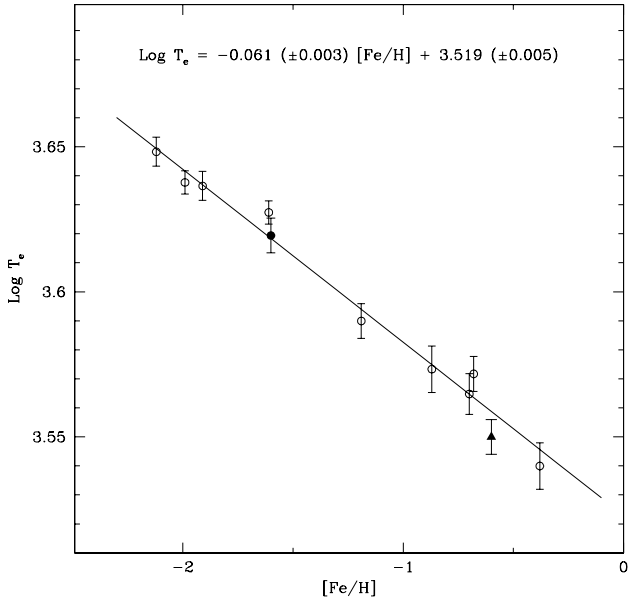}
   \caption{$\log T_e$ at $M_{Bol}=-3$ as a function of [Fe/H] for the
   ten GGCs of F00 (open circles), RGB-MP (filled circle) and the RGB-a
   (filled triangle). }
   \label{bol}%
   \end{figure}

\subsection{The Theoretical Plane}

The transformation of the RGB observed colors and magnitudes into the
theoretical plane ($M_{Bol}$, $\log T_{e}$) has been performed using
the bolometric corrections and temperature scale for Population II
giants computed by Montegriffo et al. (1998). In Figure~\ref{teo} the
fiducial lines for the two populations of $\omega$~Cen are compared to
a sample of GGCs from F00 in the theoretical plane. From this diagram,
we can easily derive the RGB effective temperature at a given
bolometric magnitude. Figure~\ref{bol} shows the effective temperature at
$M_{Bol}=-3$ as a function of metallicity for the RGB-MP, the
RGB-a and for the clusters sample of F00. We find a good consistency between the
data and the empirical relation of F00.


\section{The RGB bump}
\label{2mass}

   \begin{figure}
   \centering
     
   \includegraphics[width=8.7cm]{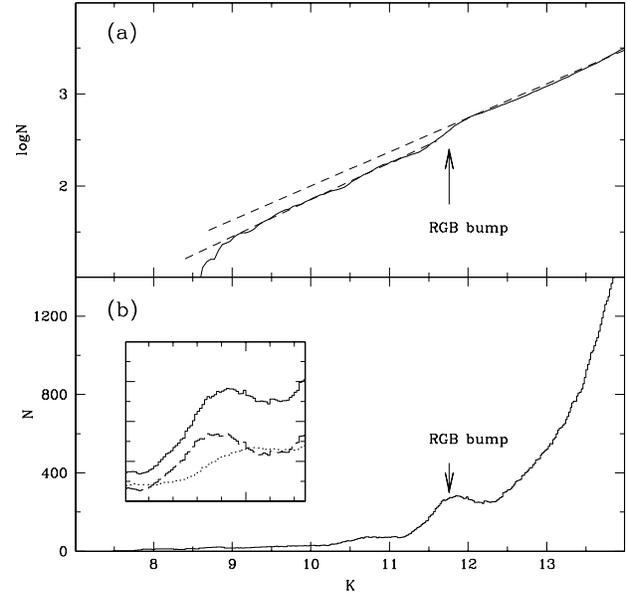}
   \caption{K-band cumulative ({\it panel} a) and differential ({\it panel} b) 
   luminosity functions of the RGB-MP, shown as smoothed histograms. The
   slope variation and the RGB bump location  have been marked.
   The contribution of the redder part of the RGB (dotted line) and of
   the bluer part (dashed line) to the bump is also shown in the small panel.}
   \label{lumfk}%
   \end{figure}

In order to enlarge our sample, both in size and in covered area, we
correlated our catalog with that obtained in the external region of the
cluster by the 2MASS survey, which 
extends to a very wide area of $3^{\circ}\times2^{\circ}$. 
Using 1,419 common stars,
we verified the absence of any significant magnitude offset
(see Section 2), hence we constructed an extended catalog where our
magnitudes were maintained in the overlapping regions, while in the
outer regions ($r>6'$) the 2MASS catalog objects were added. We thus
obtained J and K magnitudes for a global sample of more than 120,000
stars, that allowed us to identify with good accuracy one of the most
subtle features along the RGB: the RGB bump (Iben 1968). 

The identification of the RGB bump is not an easy task because of the
need of large observational samples (Crocker \& Rood 1984, Fusi Pecci
et al. 1990). Moreover, since in metal poor clusters the RGB bump
occurs at brighter luminosities, in a region that is intrinsically
poorly populated, its identification is even more difficult. To
correctly locate the bump we used both the cumulative and differential
RGB luminosity functions (LF), by detecting the slope variation in the
former and identifying the corresponding peak in the latter (Fusi Pecci
et al. 1990). We identified the RGB-MP bump with good
accuracy both in the K-band and in the V-band (using the optical
catalogue by Pancino et al. 2000) at the observed magnitudes
$V_{Bump}=14.40\pm0.05$ and $K_{Bump}=11.76\pm0.05$. The absolute
values turn out to be $M_{V}^{Bump}=0.41\pm0.14$ and 
$M_{K}^{Bump}=-1.93\pm0.14$, with the adopted distance modulus and reddening
corrections.

\begin{figure}
   \centering
     
   \includegraphics[width=8.7cm]{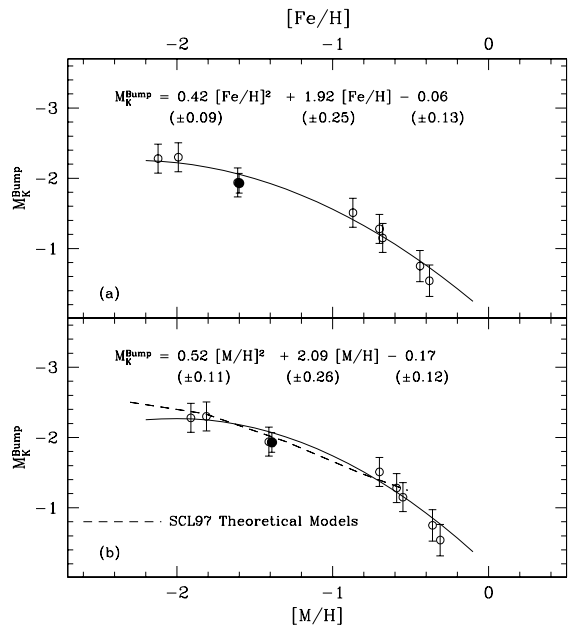}
   \caption{ $M_K$ at the RGB bump as a function of the metallicity
   [Fe/H] ({\it panel} a) and in the global scale ({\it panel} b) for 8 GGCs of
   F00 (open circles) and the RGB-MP (filled circles). The dashed line
   in {\it panel} b is the theoretical prediction by Straniero, Chieffi \&
   Limongi (1997) models at t=16Gyr.}
   \label{bump}%
   \end{figure}

Our detection of the RGB bump is illustrated in Figure~\ref{lumfk},
where the LF of the RGB-MP in the K band is shown as a smoothed
histogram. In Figure~\ref{bump} the absolute K magnitude of the bump is
compared with the empirical F00 relation and with the
theoretical models by Straniero, Chieffi \& Limongi (1997). The
location of the RGB-MP bump is in reasonable agreement with the
predictions of both relations. However, the shape of the bump on the
differential LF appears widened, as can be expected due to a possible
residual contamination of RGB-MInt stars in our RGB-MP sample. To test
this hypothesis, we have compared in Figure~\ref{lumfk} the position of
the RGB-MP bump with the bump position computed for the blue side of
the RGB-MP and the red side of the RGB-MP, separately. As expected, the
red side bump appears shifted to fainter magnitudes while the blue side
bump appears less widened and slightly shifted towards brighter
magnitudes. The presence of multiple
bumps at different magnitudes for different sub-samples of the RGB has
been shown also by Rey et al. (2003, see their Figure 10), in agreement
with what found here. 

\section{Conclusion}

We presented an extensive near IR, J and K catalog of stars in the
giant globular cluster $\omega$~Cen. More than 73,000 stars have been
measured allowing an accurate photometric characterization of the RGB.
In particular, the colors at different magnitude levels, the magnitude
at different colors, the RGB slope and the RGB bump position have been
measured, scaled to the absolute plane and compared to similar features
measured in clusters with different metallicity by F00. The agreement
with the F00 relations is quite good, and the photometric properties of
the newly discovered anomalous RGB (RGB-a) consistently reflect the
high metal content of this sub-population found by previous spectroscopic work.  

\begin{acknowledgements}

We warmly thank Paolo Montegriffo for assistance during the catalogs
cross-correlation and astrometric calibration procedure.
We also thank Katia Cunha, the Referee of our paper, for her 
precious comments and suggestions.
 The financial support of the Agenzia Spaziale Italiana and the Ministero della
Istruzione e della Ricerca Universitaria is kindly acknowledged. This
publication makes use of data product from the Two Micron All Sky
Survey, which is a joint project of the University of Massachussets and
the Infrared Processing Data Analysis Center/California Institute of
Technology, founded by the National Aeronautics, the Space
Administration and the National Science Foundation.

\end{acknowledgements}


\end{document}